# Automatic Modulation Recognition of PSK Signals with Sub-Nyquist Sampling Based on High Order Statistics

Zhengli Xing[1], Jie Zhou[1], Jiangfeng Ye[1], Jun Yan[1], Jifeng Zou[2], Lin Zou[2], Qun Wan[2]

[1] Institute of Electronic Engineering, China Academy of Engineering Physics,

Mianyang, China

[2] School of Electronic Engineering,

University of Electronic Science and Technology of China,

Chengdu, China

**Correspondence information:**

Jie Zhou,

Mailbox 919-513, Institute of Electronic Engineering, China Academy of Engineering Physics,

Mianyang, Sichuan Province, China 621900

**E-mail address: zhengli102@163.com**

**Tel.:** 0086-18683931561

# Automatic Modulation Recognition of PSK Signals with Sub-Nyquist Sampling Based on High Order Statistics

*Zhengli Xing[1], Jie Zhou[1], Jiangfeng Ye[1], Jun Yan[1], Jifeng Zou[2], Lin Zou[2], Qun Wan[2]*
[1]Institute of Electronic Engineering, China Academy of Engineering Physics, Mianyang, China
[2]University of Electronic Science and Technology of China, Chengdu, China

*Abstract*—Sampling rate required in the $N^{th}$ Power Nonlinear Transformation (NPT) method is typically much greater than Nyquist rate, which causes heavy burden for the Analog to Digital Converter (ADC). Taking advantage of the sparse property of PSK signals' spectrum under NPT, we develop the NPT method for PSK signals with Sub-Nyquist rate samples. In this paper, combined the NPT method with Compressive Sensing (CS) theory, frequency spectrum reconstruction of the $N^{th}$ power nonlinear transformation of PSK signals is presented, which can be further used for AMR and rough estimations of unknown carrier frequency and symbol rate.

*Keywords*—compressive sensing; PSK signals; modulation classification; Nth power nonlinear transformation

## I. INTRODUCTION

In cognitive radio (CR) applications, Automatic Modulation Recognition (AMR) is a basic task [1], which performs the classification of different types of modulation. As the preprocess of demodulation, AMR can be widely used for spectral monitoring and user identification in spectrum sensing. As an active research spot, various techniques, such as Wavelet/ Fourier transform, Cumulants, have been proposed [1]-[3]. However, according to the Nyquist Sampling rule, most of the typical methods require high sampling rate, which brings heavy burden for ADC, especially for wideband signals. Furthermore, the aim of AMR and carrier frequency and symbol rate estimation only extract quite little information from the enormous data.

Recently, Compressive Sensing (CS) has been introduced as a new sampling theory [4]-[6], which can just recover signals from the Sub-Nyquist rate samples. And in CR, CS has been introduced to relieve the heavy burden of ADC [9], which utilized the sparsity of wireless signals.

Currently, most applications of CS focus on reducing the sampling rate and reconstructing signals precisely, while little attention has been paid to AMR. Lim and Wakin [6] tried to estimate the $N^{th}$ power spectrum without reconstruction. Whereas, according to the Compressive Signal Processing (CSP) theory proposed by Davenport, Wakin, et al [7], [6] does not have strong anti-noise ability, and only exploit the peak of frequency, and it can just discriminate the BPSK, QPSK and 8PSK signals. Similarly, Chai proposed a new method in [8] to estimate the compressive higher order cyclostationary statistics, which can be easily affected by noise; based on Tian's work [9], Zhou and others derived some other methods [10]-[11] by reconstructing cyclic spectrum from Sub-Nyquist samples, but the calculations are quite complicated.

In general, the task of AMR consists of two main parts: feature extraction and classification. In this paper, after reconstructing the frequency spectrum of NPT signals, we extract the primary elements. Then, with the Support Vector Machine (SVM), we can implement the classification effectively.

The remainder of this paper is organized as follows. Section II analyzes the $2^{nd}$, $4^{th}$ and $8^{th}$ nonlinear transformation of PSK type signals with uniform sampling in time domain. It describes the relationship between Sub-Nyquist rate samples and the frequency spectrum of NPT signals in Section III. Section IV introduces SVM, and propose the classification strategy with primary elements of the spectrum, as well as rough estimation methods of carrier frequency and symbol rate. Simulation results are presented in Section V. Finally, conclusions are made in Section VI.

## II. FEATURES OF NPT FOR PSK TYPE SIGNALS

Determining the number of discrete peaks in frequency spectrum of signals undergone NPT is the key step. According to that, modulation type can be ascertained. Besides, exploiting the locations of the peaks, we can get rough estimations of the carrier frequency $f_c$ and symbol rate $R_s$.

Here, based on the previous work [4], we extend the analysis to 8PSK and OQPSK, and we summarize the spectrum features of NPT signals for PSK modulation.

### A. Signal Model

For MPSK (M=2, 4, 8), the signal models are as follows [11]:

$$s_{MPSK} = \sum_{n=-\infty}^{\infty} A g(t - nT_s) \exp\left( j2\pi \frac{(m_n - 1)}{M} + j2\pi f_c \right) \quad (1)$$

where $A$ is the amplitude, $T_s = 1/R_s$ is the symbol period, $M \in \{2,4,8\}$ is the number of unique phases used, $m_n$ is the $n$th transmitted symbol, $f_c$ is the carrier frequency, g($t$) is the square-root raised cosine filter (SRRC), α is the roll-off factor.

For OQPSK, the signal model is[12]:

$$s_{OQPSK} = A[I_n + j*Q_n]\exp(j2\pi f_c t) \quad (2)$$

and

$$I_n = \sum_n a_n g(t - (2n-1)T_b)$$
$$Q_n = \sum_n b_n g(t - 2nT_b) \quad (3)$$

where $a_n, b_n \in \{\pm 1\}$ are i.i.d. (independently identically distributed) random sequences. $T_b = T_s/2$ is the bit period.

For MSK, the signal model is[12]:

$$s_{MSK} = A[I_n + j*Q_n]e^{j2\pi f_c t} \quad (4)$$

and

$$I_n = \sum_n a_n \text{rect}(t - (2n-1)T_b)\cos(\frac{\pi t}{2T_b})$$
$$Q_n = \sum_n b_n \text{rect}(t - 2nT_b)\sin(\frac{\pi t}{2T_b}) \quad (5)$$

where $a_n, b_n \in \{\pm 1\}$ are i.i.d. (independently identically distributed) random sequences. $\text{rec}(\cdot)$ denotes the rectangular function.

*B. Sparsity of Signals after NPT*

As given in [4], we can extend the analysis method to the 8PSK and OQPSK signals, whose spectrum features are shown in Table 1 and Table 2. According to Table 2, the number of discrete peaks is different for PSK signals, and the peaks' locations are determined by $f_c$ and $R_s$. Thus, rough estimations of $f_c$ and $R_s$ can be fulfilled via those locations.

As an example, in Fig. 1(a), it can be clearly seen that the spectrum of BPSK signals raised to the power of 2 are sparse. There only exist three prominent discrete peaks in the spectrum, which are in fact what we care. As the spectrum is approximately sparse, it suggests that the CS theory can be applied. What should also be considered is the influence of the roll-off factor, which is discussed in Section III-C below.

III. FEATURES FROM SUB-NYQUIST SAMPLING

*A. CS Sampling*

In CS theory, high-rate uniform sampling is replaced by low-rate random sampling, which can significantly reduce the number of samples [4]-[6]. Here, from CS theory, we can design the measurement matrix in our application as Gaussian random matrices, whose coherence with a fixed orthonormal basis is very low [4]-[6].

Here, we define $\mathbf{z} \in \mathbb{C}^{L \times 1}$ to be a hypothetical vector of uniform samples of the received signal, $\mathbf{z}_N$ is the vector undergone NPT. Then, we have:

$$\mathbf{z}_N[l] = (\mathbf{z}[l])^N, \quad l = 0,1,...,L-1 \quad (6)$$

Just as shown in Table 2, PSK signals raised to a certain power are sparse in frequency domain. Thus, we choose DFT synthesis matrix as the sparsifying matrix, and the compressive sampling procedure can be written as:

$$\mathbf{y} = \mathbf{\Phi}\mathbf{z}_N = \mathbf{\Phi}\mathbf{\Psi}\mathbf{f}_N \quad (7)$$

where $\mathbf{\Phi} \in \mathbb{R}^{M \times L}$ is the measurement matrix ($M \ll L$), $\mathbf{\Psi} \in \mathbb{C}^{L \times L}$ is the DFT synthesis matrix, $\mathbf{f}_N \in \mathbb{C}^{L \times 1}$ is the vector of DFT coefficients of $\mathbf{z}_N$, $\mathbf{y}_N \in \mathbb{C}^{M \times 1}$ is the vector of Sub-Nyquist samples.

Thanks to the sparsity of $\mathbf{f}_N$, it can be recovered with CS reconstruction method.

*B. Spectrum Recostruction*

The problem of recovering $\mathbf{f}_N$ from $\mathbf{y}_N$ is an NP-hard problem. It can be converted to $l_1$-norm optimization problem in CS [15] as:

$$\hat{\mathbf{f}}_N = \arg\min \|\mathbf{f}_N\|_1 \quad s.t. \quad \mathbf{y}_N = \mathbf{\Phi}\mathbf{z}_N = \mathbf{\Phi}\mathbf{\Psi}\mathbf{f}_N \quad (8)$$

Here, we solve the convex problems by CVX [15].

In order to reduce the influence of the noises on the frequency spectrum, a piecewise smoothing procedure is used as:

$$\tilde{\mathbf{f}}_N(i) = \mathbf{f}_N(i) - \frac{1}{2l}\left(\sum_{j=1}^{l}(\mathbf{f}_N(i+j) + \mathbf{f}_N(i-j))\right) \quad (9)$$

where $\tilde{\mathbf{f}}_N$ denotes the smoothing result of $\mathbf{f}_N$, $l$ is the number of smoothing points.

The processing procedure of (9) can be rewritten as matrix operation:

$$\tilde{\mathbf{f}}_N = \mathbf{B}\mathbf{f}_N \quad (10)$$

where

$$\mathbf{B} = \begin{bmatrix} \frac{-1}{2l} & \cdots & \frac{-1}{2l} & 1 & \frac{-1}{2l} & \cdots & \frac{-1}{2l} & 0 & 0 & 0 & 0 & \cdots & 0 & 0 & 0 & 0 \\ \frac{-1}{2l} & \cdots & \frac{-1}{2l} & 1 & \frac{-1}{2l} & \cdots & \frac{-1}{2l} & 0 & 0 & 0 & 0 & \cdots & 0 & 0 & 0 & 0 \\ \vdots & \ddots & \vdots & \vdots & \vdots & \ddots & \vdots & \vdots & \vdots & \vdots & \vdots & \ddots & \vdots & \vdots & \vdots & \vdots \\ \frac{-1}{2l} & \cdots & \frac{-1}{2l} & 1 & \frac{-1}{2l} & \cdots & \frac{-1}{2l} & 0 & 0 & 0 & 0 & \cdots & 0 & 0 & 0 & 0 \\ 0 & \frac{-1}{2l} & \cdots & \frac{-1}{2l} & 1 & \frac{-1}{2l} & \cdots & \frac{-1}{2l} & 0 & 0 & 0 & \cdots & 0 & 0 & 0 & 0 \\ 0 & 0 & \frac{-1}{2l} & \cdots & \frac{-1}{2l} & 1 & \frac{-1}{2l} & \cdots & \frac{-1}{2l} & 0 & 0 & \cdots & 0 & 0 & 0 & 0 \\ 0 & 0 & 0 & 0 & \cdots & 0 & 0 & 0 & 0 & \ddots & 0 & \cdots & \vdots & \vdots & \vdots & \vdots \\ 0 & 0 & 0 & 0 & 0 & \cdots & 0 & 0 & 0 & 0 & \frac{-1}{2l} & \cdots & \frac{-1}{2l} & 1 & \frac{-1}{2l} & \frac{-1}{2l} \\ 0 & 0 & 0 & 0 & 0 & 0 & \cdots & 0 & 0 & 0 & 0 & \frac{-1}{2l} & \cdots & \frac{-1}{2l} & 1 & \frac{-1}{2l} \\ 0 & 0 & 0 & 0 & 0 & 0 & \cdots & 0 & 0 & 0 & 0 & 0 & \frac{-1}{2l} & \cdots & \frac{-1}{2l} & 1 \end{bmatrix}$$

$\leftarrow$ the $l^{th}$ row

$$(11)$$

Table 1. Peak lines in different nonlinearities for PSK type signals

| Nonlinearity | Frequency | Modulation Type | | | | |
|---|---|---|---|---|---|---|
| | | BPSK | QPSK | 8PSK | OQPSK | MSK |
| None | $f_c$ | N | N | N | N | N |
| | $f_c + kR_s$ | N | N | N | N | N |
| $(.)^2$ | $2f_c$ | Y | N | N | N | N |
| | $2f_c + (k+0.5)R_s$ | N | N | N | N | Y |
| | $2f_c + kR_s$ | Y | N | N | Y | N |
| $(.)^4$ | $4f_c$ | Y | Y | N | Y | N |
| | $4f_c + (k+0.5)R_s$ | N | N | N | N | N |
| | $4f_c + kR_s$ | Y | Y | N | Y | Y |

Table 2. Number of discrete peaks for PSK type signals

| Nonlinearity | Modulation Type | | | | |
|---|---|---|---|---|---|
| | BPSK | QPSK | 8PSK | OQPSK | MSK |
| None | 0 | 0 | 0 | 0 | 0 |
| $(.)^2$ | 3 | 0 | 0 | 2 | 2 |
| $(.)^4$ | 3(or 5) | 3(or 5) | 0 | 3(or 5) | 2 |
| $(.)^8$ | 3(or 5) | 3(or 5) | 3(or 5) | - | - |

With the smoothing procedure, the model of (8) can be adapted as:

$$\tilde{\mathbf{f}}_N = \arg\min \|\mathbf{B}\mathbf{f}_N\|_1 \quad s.t. \quad \mathbf{y}_N = \mathbf{\Phi}\mathbf{z}_N = \mathbf{\Phi}\mathbf{\Psi}\mathbf{f}_N \quad (12)$$

For example, if $N=2$, the model can be rewritten as:

$$\tilde{\mathbf{f}}_2 = \arg\min \|\mathbf{f}_2\|_1 \quad s.t. \quad \mathbf{y}_2 = \mathbf{\Phi}\mathbf{z}_2 = \mathbf{\Phi}\mathbf{\Psi}\mathbf{f}_2 \quad (13)$$
$$\mathbf{z}_2 = [z_1^2 \ldots z_L^2]^T$$

By solving the model of (12), the smoothing frequency spectrum of signals after NPT can be obtained, which reduces the impact of Gaussian white noises.

The spectrum of BPSK signal in power 2 is shown in Fig. 1(a). Fig. 1(b) is the reconstructed spectrum by CVX. It is obvious that the three major discrete peaks can be completely recovered.

*The Roll-off Factor Effect*

Here, for analytical simplicity, we define the energy ratio $r_p$ as the ratio of the peaks' energy to the entire signal, which measures whether the peaks are high enough to be used and for CS reconstruction methods to be applied:

$$r_p = E_p / E_s \quad (14)$$

where $E_p$ is the energy of the discrete peak's energy, and $E_s$ is the entire signal's energy.

In our simulation, the BPSK signal contains 2048 symbols, and has $f_c = 0.5kHz$, $R_s = 0.8kHz$, $f_s = 6.4kHz$, and the roll-off ranges in (0, 1).

The $r_p$ corresponding to $2f_c$ and $2f_c + R_s$ of BPSK signals raised to the power of 2 are shown in Fig. 2. As shown in Fig. 2, $r_p$ increases as the roll-off factor rises. When α is too small, the ratio of $2f_c + R_s$ will become too small for AMR, and the reconstruction can not work either.

IV. CLASSIFICATION AND ESTIMATION

Since the vector of spectrum is of high dimension, we use the Support Vector Machine (SVM) to implement AMR efficiently.

*A. SVM*

SVM can be used both for linear and nonlinear classifications with corresponding kernel function. Supposing the original data is $x = [x_1, \cdots, x_L]^T$, and the mapping function is $\Phi(x_i)$, then we are able to map the original data into a high-dimensional feature space, where the classification can be made.

The function $K(x_i, x_j) = \Phi(x_i) \cdot \Phi(x_j)$ is called the kernel function. In SVM theory, there are various kernel functions, such as linear, polynomial and Gaussian RBF, et al, which are shown in Table 3. Thanks to its perfect performance, RBF function is used in our simulation.

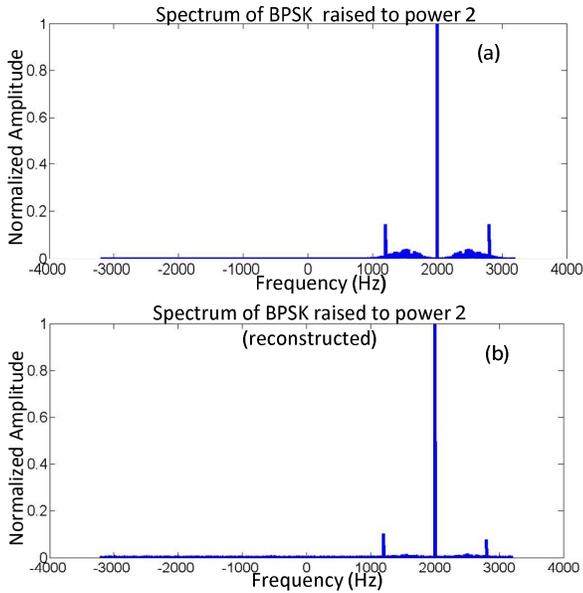

Fig.1. Simulation results for BPSK signals after NPT

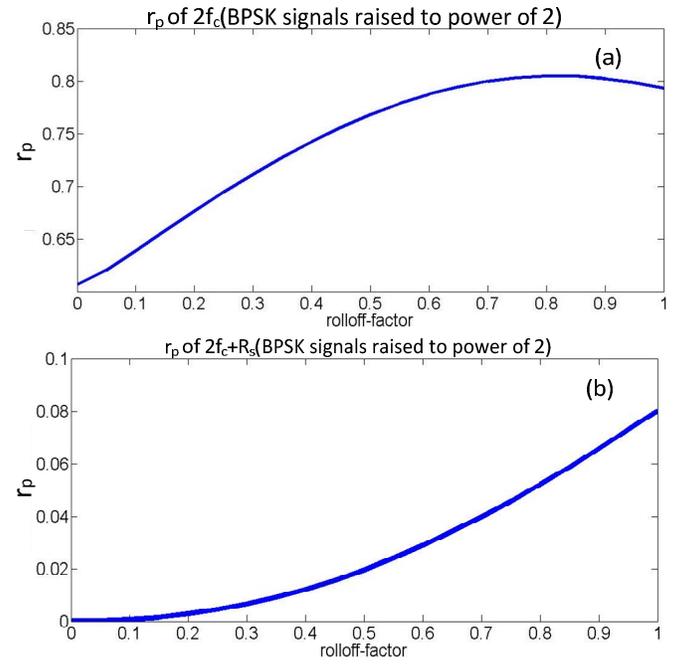

Fig.2. The effect of roll-off factor on energy ratio of peaks

Table 3. Formulation of common used kernel functions

| Kernel | $K(x_i, x_j)$ |
|---|---|
| Linear | $x_i^T \cdot x_j$ |
| Polynomial | $(\gamma x_i^T \cdot x_j + r)^d$ |
| Gaussian RBF | $\exp(-\|x_i - x_j\|^2 / 2\sigma^2)$ |

However, as we all know, practical classification problems always contain more than two classes. A common way to construct a k-class SVM is to combine several Binary classifers together. Some popular methods, such as 'one-against-one', 'one-against-all' and 'DDAGSVM', can be used to solve the k-class problem [19].

In our simulation below, we use the SVM toolbox (libsvm), provided by Chih [20], which uses the 'one-against-one' strategy.

### B. AMR Strategy

As shown in Fig. 1, the energy of the signals are intensively distributed in those discrete peaks. Furthermore, in order to decrease the computation complexity, we need to reduce the dimension of the data. There are many classical methods to realize the purpose, such as PCA, MDS, et al [21]. Here, we propose a simple method to extract the main elements of the spectrum instead of those high complexity methods.

From Table 2, there are at most 5 discrete peaks of the spectrum, and what we concerns is the energy of the peaks relative to other part. That means, instead of using the whole vector of spectrum, we only need to retain the $m$ largest elements of the spectrum, where $m$ is a preset parameter. In our simulation below, we set $m = 20$.

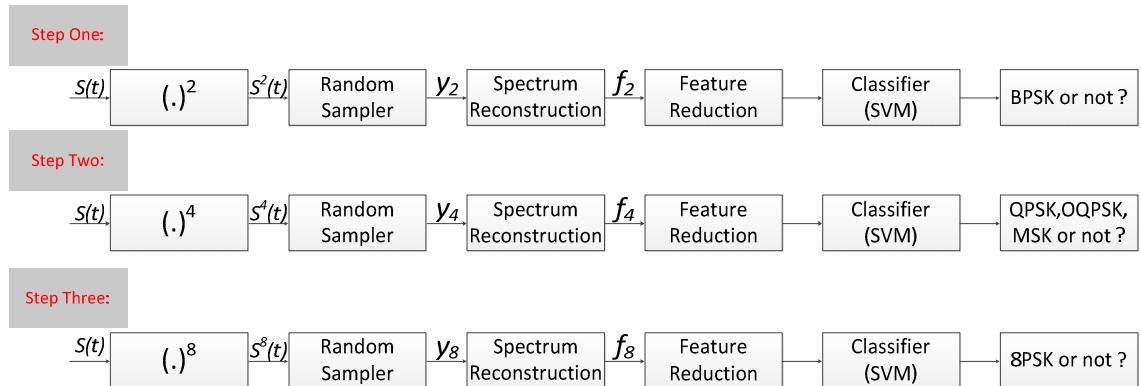

Fig.3. The proposed AMR scheme

Besides, to distinguish BPSK signals, we only need $f_2$. With $f_2$ and $f_4$, we can distinguish QPSK, OQPSK and MSK. As for 8PSK, it then make full use of $f_2$, $f_4$ and $f_8$.

The proposed AMR framework is shown in Fig. 4. We use a random sampler at a Sub-Nyquist rate to sample the signals undergone NPT, feature reduction is performed after frequency spectrum reconstruction and the lower-dimensional feature is sent to SVM, which to make the final decision as a classifier.

*C. Estimation of $f_c$ and $R_s$*

As given in Table 1, for each kind of signal, the locations of discrete peaks are determined by $f_c$ and $R_s$ except 8PSK. The locations of discrete peaks can be used to roughly estimate $f_c$ and $R_s$. Here, we take estimation methods of $f_c$ and $R_s$ for QPSK for instance, and other kinds of signals can be obtained in a similar way.

Here, denote $A_1$, $A_2$, $A_3$ as the locations of three dominant peaks of $f_4$ of QPSK signal, and it can be calculated that:

$$\hat{f}_c = A_1/4 \left(\text{or } (A_2 + A_3)/8\right) \quad (15)$$

$$\hat{R}_s = |A_1 - A_2|/2 \quad (16)$$

## V. SIMULATIONS

The proposed methods are tested and verified in this section. Simulation scenarios are set as follows: symbol number is 1024, $\alpha = 0.5$, $f_c = 0.5 kHz$, $R_s = 0.8 kHz$. Here, the Nyquist-rate of uniform sampling is $f_s = 6.4 kHz$ to avoid aliasing, which means $L = 8192$. And the uniform samples corresponding to the "Nyquist rate" curves in Fig. 5. Besides, the number of Sub-Nyquist rate samples is $M$. Here, we can define the compression ratio as:

$$\beta = M/L \quad (17)$$

here, we set $\beta = 0.3$.

Here, in Fig. 5(a) and Fig. 5(b), we define $r_\alpha$ as the rate of correct classification.

Fig. 5(a) and Fig. 5(b) depict the rate of correct classifications versus varying signal-to-noise ratios (SNR). Fig. 5(c) shows the estimation results of carrier frequency of QPSK signal. From Fig. 5(a) and Fig. 5(b) is that, for a given accuracy rate of classification, AMR using Sub-Nyquist rate requires about 2 dB (5dB for BPSK) of SNR more than uniform sampling. From Fig. 5(c), it is also about 2dB for the estimation accuracy of carrier frequency.

As can be seen from the Figures, when SNR is low, the proposed method doesn't work as well as the "Nyquist rate" method. This is because when the noise' energy is high, and the spectrum is no longer sparse. Thus, the CS theory also doesn't apply to this condition.

As discussed above, while reducing the computation amounts, the method proposed by [6] can just discriminate the BPSK, QPSK and 8PSK, while the method we proposed here extend the analysis to OQPSK and MSK. Some other signals we doesn't mention here, such as FSK and ASK, whose spectrums are distinguishable [3] can also use the method we proposed to classify.

What's more, as the statistical characters we use here concern much about the symbol number. [6] used 5000 symbols in the simulation, while in this paper, we only use 1024 symbols to finish the same task, and the SNR we need for the same rate of correct classification is only much lower.

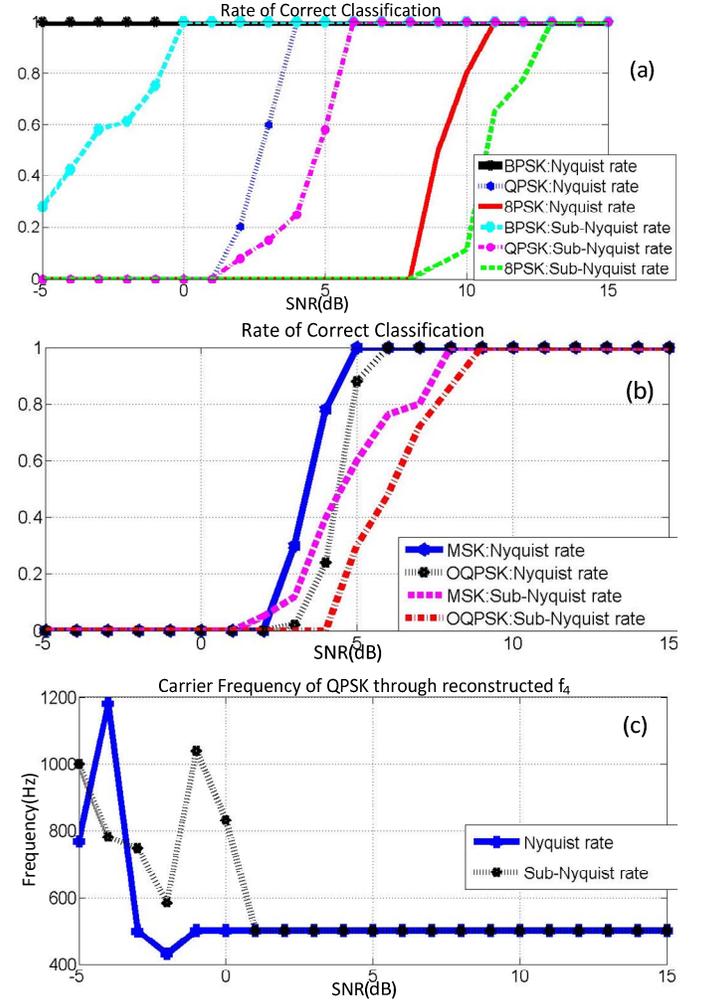

Fig.4. Simulation results of AMR and estimation of $f_c$

## VI. CONCLUSION

Combined CS with NPT method, we fulfill AMR and rough estimations of carrier frequency and symbol rate. Simulation results show the effectiveness of the proposed method.

VII. ACKNOWLEDGEMENTS

This work was supported in part by the National Nature Science Foundation of China under Grant No. 61301267, and in part by the Fundamental Research Funds for the Central Universities.

REFERENCES

[1] J. Mitola and Jr. G. Q. Maguire, "Cognitive radio: making software radios more personal," *Personal Communications, IEEE*, vol. 6, pp. 13-18, 1999.

[2] O.A. Dobre, A. Abdi, Y. Bar-Ness, "Survey of automatic modulation classification techniques: classical approaches and new trends," *IET Commun*, pp.137–156, 2007.

[3] J. Reichert, "Automatic classification of communication signals using higher order statistics," in *Proc. IEEE Int. Conf. Acoustics, Speech, and Signal Processing (ICASSP)*. San Francisco, California, USA. 1992.

[4] E. J. Candes and M. B. Wakin, "An Introduction To Compressive Sampling," *Signal Processing Magazine, IEEE,* vol. 25, pp. 21-30, 2008.

[5] M. F. Duarte, M. A. Davenport, M. B. Wakin, and R. G. Baraniuk, "Sparse Signal Detection from Incoherent Projections," in *Proc. IEEE Int. Conf. Acoustics, Speech and Signal Processing, (ICASSP),* Toulouse, France. 2006.

[6] C. W. Lim and M. B. Wakin, "Automatic modulation recognition for spectrum sensing using nonuniform compressive samples," in *Proc. IEEE Int. Conf. International Conference on Communications* (*ICC*), Ottawa Canada. 2012.

[7] M. A. Davenport, P. T. Boufounos, M. B. Wakin, and R. G. Baraniuk, "Signal Processing With Compressive Measurements," *Selected Topics in Signal Processing,* vol. 4, pp. 445-460, 2010.

[8] C. W. Lim and M. B. Wakin, "CHOCS: A Framework for Estimating Compressive Higher Order Cyclostationary Statistics," in *Proc. IEEE Int. Conf. SPIE Defense, Security, and Sensing (DSS)*. Baltimore, Maryland, USA. 2012.

[9] T. Zhi, "Cyclic Feature Based Wideband Spectrum Sensing Using Compressive Sampling," in *Proc. IEEE Int. Conf. International Conference on Communication* (ICC), Kyoto, Japan. 2011.

[10] Z. Lei and M. Hong, "Wavelet Cyclic Feature Based Automatic Modulation Recognition Using Nonuniform Compressive Samples," *Vehicular Technology Conference (VTC Fall),* pp. 1-6, 2013.

[11] Z. Lingchen, L. Chenchi and J. H. McClellan, "Cyclostationarity-based wideband spectrum sensing using random sampling," in *Proc. IEEE Int. Conf. Vehicular Technology Conference (VTC Fall)*, Las Vegas, USA. 2013.

[12] Liang Hong; Ho, K.C., "Modified CRLB on the modulation parameters of OQPSK signal and MSK signal," in *Proc. IEEE Int. Conf.* Wireless Communications and Networking Confernce. Chicago, IL, USA. 2000.

[13] E. J. Candes and T. Tao, "Near-Optimal Signal Recovery From Random Projections: Universal Encoding Strategies?" *Information Theory,* vol. 52, pp. 5406-5425, 2006.

[14] M. RUDELSON and R. VERSHYNIN, "On sparse reconstruction from Fourier and Gaussian measurements," *Information Sciences and Systems,* pp. 22-24 March 2006.

[15] Candes E J, Romberg J, Tao T. "Stable signal recovery from incomplete and inaccurate measurements". *Comm Pure Appl Math*, 59: 1207-1223, 2006

[16] I. CYX Research, "CYX: Matlab software for disciplined convex programming, version 2.0 beta." Software available at: http://cvxr.com/cvx , Sept. 2012

[17] V. Vapnik, "The Nature of Statistical Learning Theory". *Springer*, 1995.

[18] C. Latry, C. Panem and P. Dejean, "Cloud detection with SVM technique," in *Proc. IEEE Int. Conf. Geoscience and Remote Sensing Symposium (IGARSS).* Barcelona, Spain, 2007.

[19] L. Hua and X. Z. Yong, "An algorithm of soft fault diagnosis for analog circuit based on the optimized SVM by GA," in *Proc. IEEE Int. Conf. Electronic Measurement & Instruments (ICEMI)*, Beijing, China, 2009.

[20] Chih-Chung Chang and Chih-Jen Lin, LIBSVM： a library for support vector machines, 2001. Software available at: http://www.csie.ntu.edu.tw/~cjlin/libsvm

[21] Chauhan, S.; Prema, K.V., "Effect of dimensionality reduction on performance in artificial neural network for user authentication," in *Proc. IEEE Int. Advance Computing Conference (IACC),* Ghaziabad, India. 2013